\documentclass[10pt, aps, prapplied, twocolumn, showpacs, floatfix, superscriptaddress, showkeys]{revtex4-2}
\usepackage{amsmath, amssymb, mathtools, array, color, bbm}
\usepackage{hyperref}
\hypersetup{colorlinks=true, linkcolor=blue, citecolor=blue}
\usepackage{graphicx}
\usepackage{times}
\usepackage{tabularx}
\usepackage{booktabs}

\begin{document}


\title{Transformation-Dependent Performance-Enhancement of Digital Annealer for 3-SAT}

\author{Christian~M{\"u}nch}
\affiliation{Fujitsu Services GmbH, Mies-van-der-Rohe-Stra{\ss}e 8, 80807 Munich, Germany}
\author{Fritz Schinkel}
\affiliation{Fujitsu Services GmbH, Mies-van-der-Rohe-Stra{\ss}e 8, 80807 Munich, Germany}
\author{Sebastian~Zielinski}
\affiliation{Institute for Informatics, LMU Munich, Oettingenstra{\ss}e 67, 80538 Munich, Germany}
\author{Stefan~Walter}
\affiliation{Fujitsu Services GmbH, Mies-van-der-Rohe-Stra{\ss}e 8, 80807 Munich, Germany}

\date{\today}


\begin{abstract}
Quadratic Unconstrained Binary Optimization (QUBO) problems are NP-hard problems and many real-world problems can be formulated as QUBO. Currently there are no algorithms known that can solve arbitrary instances of NP-hard problems efficiently. Therefore special-purpose hardware such as Digital Annealer, other Ising machines, as well as quantum annealers might lead to benefits in solving such problems. We study a particularly hard class of problems which can be formulated as QUBOs, namely Boolean satisfiability (SAT) problems, and specifically 3-SAT. One intriguing aspect about 3-SAT problems is that there are different transformations from 3-SAT to QUBO. We study the transformations' influence on the problem solution, using Digital Annealer as a special-purpose solver. Besides well-known transformations we investigate a novel in this context not yet discussed transformation, using less auxiliary variables and leading to very good performance. Using exact diagonalization, we explain the differences in performance originating from the different transformations. We envision that this knowledge allows for specifically engineering transformations that improve a solvers capacity to find high quality solutions. Furthermore, we show that the Digital Annealer outperforms a quantum annealer in solving hard 3-SAT instances.

\end{abstract}

\maketitle

\section{Introduction}\label{sec:intro}
Combinatorial optimization deals with discrete decisions and finding the best combination in a typically huge search space~\cite{Wolsey1999, Papadimitriou2013}. Such problems are ubiquitous.
Academic problems such as MaxCut, set partitioning, or graph coloring, as well as many practically relevant problems like task scheduling, vehicle routing,
or knapsack are prime examples of combinatorial optimization problems.
Combinatorial optimization problems have the unpleasant characteristic in common that the search space grows rapidly (i.e. exponentially or factorially) with the number of decision variables
(i.e., possible yes-no decisions). This is the reason why an exhaustive
search approach is very inefficient with increasing problem size. Therefore, many heuristics, such as simulated annealing~\cite{Kirkpatrick1985}, tabu search~\cite{Glover1986}, and others are employed to solve such problems.

A promising framework to solve combinatorial optimization problems is the Quadratic Unconstrained Binary Optimization (QUBO) approach~\cite{Lu2010, Kochenberger2014, Glover2019}.
QUBO problems deal with minimizing a polynomial of degree two with 0 and 1 valued variables. Since there is a correspondence of the QUBO
formulation to the well-known and well-studied Ising spin glass problem~\cite{Barahona1988, Lucas2014, Boettcher2019}, the QUBO formulation gained popularity with the availability of Quantum Annealing hardware~\cite{dwave1, dwave2, dwave3, dwave4, dwave5, dwave6} which was quickly followed by other types of Ising machines, such as Fujitsu's Digital Annealer~\cite{da1, da2, da3}, Toshiba's Simulated Bifurcation machine~\cite{toshiba}, and others~\cite{isingmachines}.

Besides the aforementioned combinatorial optimization problems, Boolean satisfiability (SAT) problems (especially 3-SAT) problems are an important problem class, e.g., to study computational complexity and for instance showing NP-completeness~\cite{Cook1971, Levin1973}. SAT problems being constraint satisfaction problems, are of great practical relevance, with applications mainly in software verification \cite{Franjo2005, ivanvcic2008efficient} and hardware verification \cite{Grimm2018}. There are various algorithmic approaches on classical CPU and GPU hardware to solve large and complex SAT problems \cite{Gong2017survey}. However, up to now there is no algorithm known which is able to solve arbitrary SAT/3-SAT instances in (worst-case) polynomial time.

3-SAT problems, being combinatorial problems, can also be formulated as QUBO and novel compute techniques which could provide an advantage over traditional technologies can be used. In fact, there are studies on (Max-)3-SAT \cite{Choi2010, Chancellor2016, Zielinski2023Influence, Nuesslein2023}, 3XOR-SAT~\cite{Kowalsky2022}, and other variants of satisfiability problems using quantum annealers, Digital Annealers, and other promising novel compute architectures which investigate the solvers' abilities.

In this paper, we use Fujitsu's Digital Annealer to solve 3-SAT problems and investigate the performance of different 3-SAT to QUBO
transformations. We explain the differences using small problem instances which allow for exact diagonalization, and show that extrapolating the findings to larger instances can be done. We also show that Digital Annealer outperforms a quantum annealer on solving hard 3-SAT instances.

The paper is structured as follows. In Sec.~\ref{sec:background} we formally introduce the 3-SAT problem, review existing 3-SAT to QUBO transformations, introduce one novel, in this context not yet discussed, transformation, and briefly present the main features of Fujitsu's Digital Annealer. We continue in Sec.~\ref{sec:results} with our results obtained using the Digital Annealer. In Sec.~\ref{sec:analysis} we present the analysis of the results. Finally, we conclude and give an outlook in Sec.~\ref{sec:conclusion}.

\section{Background}\label{sec:background}

\subsection{The 3-SAT Problem and its Relation to QUBO}\label{sec:3SATQUBO}
In general, Boolean satisfiability problems deal with the assignment of Boolean variables $x_{0}, x_{1}, ..., x_{n-1}$ with $x_{i} \in \{0,1\}$ for $n\in \mathbb{N}$ and $i=0,\ldots,n-1$ such that
a Boolean formula $\Phi$  is satisfied, i.e., $\Phi(x_{0}, x_{1}, ..., x_{n-1})=1$. A Boolean formula $\Phi$ consists of literals (i.e., Boolean variables)
and clauses. Literals can be positive (i.e., they are represented by the variable itself) or negative (i.e., they are represented by the negated ${\rm{NOT}}$  ($\neg$)) variable. Literals are connected by the logical ${\rm{OR}}$ ($\vee$) operator and clauses are joined by the logical ${\rm{AND}}$
($\wedge$) operator. By definition a Boolean formula is in conjunctive normal form (CNF) if it is a conjunction of clauses. Here, we focus on 3-SAT
problems in CNF, i.e., clauses contain exactly three variables. An example with $m=3$ clauses and $n=3$ variables is for instance
\begin{align}\label{eqn:3sat_ex}
	\Phi_{0} = 	&(x_{1} \vee x_{2} \vee \neg x_{3}) \nonumber \\ 
			\wedge &(\neg x_{1} \vee x_{2} \vee x_{3}) \nonumber\\
			\wedge &(\neg x_{1} \vee x_{2} \vee x_{3}) \nonumber\\
			\wedge &(x_{1} \vee \neg x_{4} \vee x_{5}) \, , 
\end{align}
with the assignment $x_{1}=0, x_{2}=1, x_{3}=1, x_{4}=1, x_{5}=1$ satisfying $\Phi_{0}$. 3-SAT problems have been proven to be
NP-complete~\cite{Cook1971, Levin1973} which makes certain instances extremely hard to solve with increasing problem size. For 3-SAT, in Ref.~\cite{Gent1995} it was shown that the ratio of clauses to literals $m/n$ is a good indicator for the hardness of the particular instance. In particular, 3-SAT instances generated by randomly
drawing literals from a uniform distribution are potentially hard for $m/n \approx 4.24$~\cite{Gent1995}.
For a thorough introduction to satisfiability problems we refer to Ref.~\cite{Biere2021}.

As mentioned in Sec.~\ref{sec:intro}, combinatorial optimization problems in the form of QUBOs could potentially be solved efficiently on quantum annealers, gate-based quantum computers (via the variational Quantum Approximate Optimization Algorithm), and on other types of Ising machines such as Fujitsu's Digital Annealer. The general form of a QUBO is
\begin{align}\label{eqn:qubo}
	H(\vec{x}) = \sum_{i=0}^{n-1} Q_{ii} x_{i} + \sum_{i=0}^{n-2}\sum_{j=i+1}^{n-1} Q_{ij} x_{i} x_{j}\, ,
\end{align}
where $n\in \mathbb{N}, n\geq 2$ defines the problem size and $Q_{ij} \in \mathbb{R}$ for $i,j \in \{0,\ldots, n-1\}$ are the entries of an $n \times n$ upper triangular matrix ${\bf{Q}}$. Moreover, $\vec{x}$ is a vector of binary variables $x_{i} \in \{0, 1\}$ for $i=0,\ldots, n-1$.
The QUBO optimization problem is, to find a vector of binary values for the decision variables $x_{i}$ such that $\vec{x}$ minimizes $H(\vec{x})$.
A mapping from binary variables $x_{i}$ to spin variables $s_{i} \in \{-1, 1\}$ via $s_{i} \mapsto x_{i} = (s_{i} + 1)/2$ for $i=0,\ldots, n-1$, transforms the QUBO to the well-known Ising spin glass problem. Therefore, finding the ground state of the Ising Hamiltonian $\tilde{H}(\vec{s}) = \sum_{i=0}^{n-1} h_{i} s_{i} + \sum_{i=0}^{n-2}\sum_{j=i+1}^{n-1} J_{ij} s_{i} s_{j}$ (the local fields $h_{i}$ and the couplings $J_{ij}$ define the Ising problem) is equivalent of solving the QUBO problem~\cite{{Lucas2014, Boettcher2019}}. 

The transformation from 3-SAT to QUBO form has already been studied in literature. There are various different transformations available:
Choi has introduced a reformulation of 3-SAT instances to instances of a Maximum Weight Independent Set problem which can then easily be expressed in QUBO form~\cite{Choi2010}. Chancellor et al. have shown how to express 3-SAT instances as Ising spin glass models \cite{Chancellor2016}.
In Ref.~\cite{Nuesslein2023}, N{\"u}{\ss}lein et al. have proposed a method making use of patterns of the 3-SAT problem to algorithmically construct QUBOs from 3-SAT instances. The approach in Ref.~\cite{Nuesslein2023} has been further extended to create thousands of new 3SAT-to-QUBO transformations algorithmically (i.e. by an automatic procedure, not manually) by Zielinski et al. in Ref.~\cite{Zielinski2023}. In this work, we introduce another, previously not discussed transformation which we call CountTrue.

For the remainder of the work, we focus on Chancellor, pattern QUBO transformations, and CountTrue transformations because they show the best results in terms of solution quality. In the next section we briefly review the Chancellor, the algorithmic method, to produce thousands of 3SAT-to-QUBO transformations automatically, and introduce the CountTrue transformation.

\subsection{Chancellor Transformation}\label{sec:Chancelor}
Chancellor's transformation~\cite{Chancellor2016} starts with the binary representation of a 3-SAT formulas' clause and constructs a sum of parameter
dependent QUBOs representing the satisfiability problem for the formula as a QUBO minimization problem. In particular, given a binary representation of a
clause $I(\vec{l}, \vec{x}) = - (l_1 x_1 \vee l_2 x_2 \vee l_3 x_3)$ with $x_1, x_2, x_3 \in \{0,1\}$ and $l_1, l_2, l_3 \in \{\neg \neg\ , \neg\}$ expressing
the not-negation or negation of the literals, the clause is then transferred to Ising form. Specifically, for $i=1,2,3$ one transforms
$x_i \mapsto s_i = \left(2 x_i - 1\right)\in \{-1,1\}$ and
\begin{align}\label{eqn:chan0}
	l_i\mapsto c_i = \begin{cases}
	  			1, & \text{if }  l_i = \neg\neg \\
	  			-1, & \text{if } l_i = \neg
\end{cases} \, .
\end{align}
This allows to formulate the clause $I(\vec{l}, \vec{x})$ in cubic form 
\begin{align}\label{eqn:chan1}
	I(\vec{l}, \vec{x}) \mapsto & \frac{1}{8} \tilde{I}_{\rm{clause}}(\vec{c}, \vec{s}) \nonumber \\ 
					= & - \frac{7}{8} - \frac{1}{8} \left(c_1 s_{1} + c_2 s_{2} + c_3 s_{3}\right) \nonumber \\
						&+ \frac{1}{8} \left(c_1 c_2 s_{1}s_{2} + c_1 c_3 s_{1}s_{3} + c_2 c_3 s_{2}s_{3}\right) \nonumber \\
						& - \frac{1}{8} c_1 c_2 c_3 s_{1} s_{2} s_{3} \, .
\end{align}
Equation~(\ref{eqn:chan1}) has minimal energy for each satisfying assignment and higher energy for the only non-satisfying assignment.
An auxiliary variable $s_a\in \{-1,1\}$ is introduced to reduce the cubic term to quadratic form and obtain an Ising representation of the clause.
In contrast to the proposed replacement of the cubic term $- c_1 c_2 c_3 s_{1} s_{2} s_{3}$ in Ref.~\cite{Chancellor2016} we choose the replacement
\begin{align}\label{eqn:chan2}
	 -c_1 c_2 c_3 s_{1} s_{2} s_{3} \mapsto & I_{\rm{cubic}}(\vec{s},s_a) \nonumber \\
	 						\coloneqq & J \sum_{i=1}^3\sum_{j=1}^{i-1} s_{i}s_{j} - c_1 c_2 c_3\sum_{i=1}^{3} s_{i} \nonumber \\
							 &+  2J \sum_{i=1}^{3} s_{i}s_{a} - 2 c_1 c_2 c_3 s_{a}  \, ,
\end{align}
where $J \geq 1$ is a free parameter. Although not explicitly mentioned in Ref.~\cite{Zielinski2023}, Eq.~(\ref{eqn:chan2}) is the replacement used in their experiments.
Note that with given fixed values for $s_1^*,s_2^*,s_3^*$ and free variable $s_a$, minimizing $I_{\rm{cubic}}(s_1^*,s_2^*,s_3^*,s_a)$ over
$s_a$ is a parity check for the variables $c_1 s_1^*, c_2 s_2^*, c_3 s_3^*$ (the $c_i$ are fixed by the actual clause). As a result of the minimization over $s_{a}$, the minimal energy is $E_{\rm{min}} < 0$
for uneven numbers of ones $(+1)$, i.e., $\{c_1 s_1^*, c_2 s_2^*, c_3 s_3^*\} \in \{ \{-1,-1,1\}, \{1,1,1\} \}$, and $E_{\rm{min}} + 2$ for even numbers of
ones, i.e., $\{c_1 s_1^*, c_2 s_2^*, c_3 s_3^*\} \in \{ \{-1,-1,-1\}, \{-1,1,1\} \}$.
The remaining terms in Eq.~(\ref{eqn:chan1}) imply a minimal energy $-7 + E_{\rm{min}} +2 + 6 = E_{\rm{min}} + 1$ for the non-satisfying choice
$(c_1 s_1^*, c_2 s_2^*, c_3 s_3^*) = (-1,-1,-1)$ and a minimal energy $-7 + E_{\rm{min}}$ for all satisfying choices. This results in the representation
of the clause satisfiability problem as minimization problem of
\begin{align}\label{eqn:chan3}
	I_{\rm{clause}}(\vec{s},s_a) = &- 7 - \left(c_1 s_{1} + c_2 s_{2} + c_3 s_{3}\right) \nonumber \\
						&+ \left(c_1 c_2 s_{1}s_{2} + c_1 c_3 s_{1}s_{3} + c_2 c_3 s_{2}s_{3}\right) \nonumber \\
						& + I_{\rm{cubic}}(\vec{s},s_a) \, .
\end{align}
Transforming to binary variables $s_i \mapsto x_i = (s_i + 1)/2$ for $i\in \{1,2,3,a\}$ we obtain a QUBO representing the clause satisfiability problem.
Adding up the corresponding QUBOs for each clause (with separate auxiliary variable for each clause) we obtain a $J$-dependent QUBO representing
the satisfiability problem for the formula.

In the experiments of this paper we choose the free parameter as $J=1$ and $J=5$ and call the resulting transformations ChancellorJ1 and ChancellorJ5, respectively.

\subsection{Automated Pattern QUBOs}\label{sec:AlgorithmQUBO}
In Ref.~\cite{Zielinski2023} a method for automatically creating new QUBOs, exploiting properties of Pattern QUBOs is presented. To understand this approach we first need to observe, that there are only 4 types of clauses that can occur in a 3SAT problem:
\begin{itemize}
	\item Type 0 := $(a\vee b\vee c) $
	\item Type 1 := $(a\vee b\vee \neg c) $
	\item Type 2 := $(a\vee \neg b\vee \neg c) $
	\item Type 3 := $(\neg a\vee \neg b\vee \neg c).$
\end{itemize} 
These types of clauses are created by reordering a given 3-SAT clause, such that all negated literals are at the end of the clause. Now, all that is left to create a (new) 3-SAT to QUBO transformation is to create a QUBO for each of the 4 types of clauses. That is, a 3-SAT to QUBO transformation in this approach consists of exactly 4 reusable QUBOs, which are called Pattern QUBOs - one for each type of clause. The ground states of Pattern QUBOs should correspond to correct solutions of a clause. As there are 3 bits in a clause, there are $2^3 = 8$ possible bit strings per clause, of which 7 are satisfying and only 1 is not satisfying the clause. Therefore, Pattern QUBO that corresponds to a given clause type should have at least one ground state for each of the satisfying assignments and the not satisfying assignment should correspond to an excited state. After finding a Pattern QUBO for each of the clause types one proceeds to create a QUBO for the whole 3-SAT problem. This is done, by iterating over all the clauses from the 3-SAT problem and transforming all the clauses of the problem into QUBOs (using one of the 4 previously created Pattern QUBOs) and then combining all these clause QUBOs into a single large QUBO by adding them up using the technique of superimposing. This way, a ground state in the large QUBO directly corresponds to a solution of the given 3-SAT problem.

This approach begs an immediate question: As each 3-SAT to QUBO transformation consists of 4 Pattern QUBOs, how can we find these Pattern QUBOs? In Ref.~\cite{Zielinski2023} the authors present a simple algorithmic method that solves this problem. To be able to create a Pattern QUBO for a given 3-SAT clause that satisfies the constraint that all satisfying assignments of a 3-SAT clause need to correspond to at least one ground state in the Pattern QUBO, the Pattern QUBO needs to be of dimension 4 (i.e., a $4 \times 4$ upper triangular matrix. See Ref.~\cite{Zielinski2023} for more details). Since each Pattern QUBO is a $4 \times 4$ upper triangular QUBO matrix, there are only 10 fields of values to be filled in. This is small enough for an exhaustive search algorithm to find significant amounts of solutions. Thus, given a set of values to choose from, the algorithm inserts all possible combinations of the given set of values into the 10 fields of the $4 \times 4$ Pattern QUBO matrix, to find Pattern QUBOs that correspond to one of the 4 clause types (i.e. that has at least one ground state for each satisfying assignment in the clause). Note that the Chancellor transformation can also be created automatically using this approach. Thus, because ChancellorJ1 and ChancellorJ5 behave very differently (see Ref. \cite{Zielinski2023Influence}), the performance of QUBO transformations created by this approach heavily depends on the choice of Pattern QUBOs used for the different types of clauses.

As QUBOs generated by this approach can lead to better results, when solving 3-SAT problems on quantum annealers (as shown in Ref.~\cite{Zielinski2023}), we also used the best performing QUBO of Ref.~\cite{Zielinski2023}. We will call this QUBO the \emph{AlgorihmQUBO} for the remainder of this paper. The AlgorithmQUBO consists of four Pattern QUBOs shown in Tab.~\ref{tab:patterns}
\begin{table}[!ht]
    \begin{minipage}{.475\columnwidth}\centering
    \begin{tabular}{|p{0.65cm}||p{0.65cm}|p{0.65cm}|p{0.65cm}|p{0.65cm}|}
    \hline
    & a & b & c & K \\
    \hline
    \hline
    a & & 1 & & -1 \\
    \hline
    b & & & & -1 \\
    \hline
    c & & & -1 & 1 \\
    \hline
    K & & & &  \\
    \hline
    \end{tabular}
    \vspace{.1cm}
    Type 0 - Pattern QUBO
    \end{minipage}
    \begin{minipage}{.475\columnwidth}
    \begin{tabular}{|p{0.65cm}||p{0.65cm}|p{0.65cm}|p{0.65cm}|p{0.65cm}|}
    \hline
    & a & b & c & K \\
    \hline
    \hline
    a & &  & & -1 \\
    \hline
    b & & & -1 & 1 \\
    \hline
    c & & & 1 & -1 \\
    \hline
    K & & & & 1 \\
    \hline
    \end{tabular}
    \vspace{.1cm}
    Type 1 - Pattern QUBO
    \end{minipage} 
\vspace{.25cm}
    \begin{minipage}{.475\columnwidth}
    \begin{tabular}{|p{0.65cm}||p{0.65cm}|p{0.65cm}|p{0.65cm}|p{0.65cm}|}
    \hline
    & a & b & c & K\\
    \hline
    \hline
    a &  & -1 & & 1 \\
    \hline
    b & &1 & & -1 \\
    \hline
    c & & &  & 1 \\
    \hline
    K & & & & \\
    \hline
    \end{tabular}
    \vspace{.1cm}
    Type 2 - Pattern QUBO
    \end{minipage}
    \begin{minipage}{.475\columnwidth}
    \begin{tabular}{|p{0.65cm}||p{0.65cm}|p{0.65cm}|p{0.65cm}|p{0.65cm}|}
    \hline
    & a & b & c & K \\
    \hline
    \hline
    a &  &  &  & 1 \\
    \hline
    b & &  & 1 & -1 \\
    \hline
    c & & &  & -1 \\
    \hline
     K & & & & 1 \\
    \hline
    \end{tabular}
    \vspace{.1cm}
    Type 3 - Pattern QUBO
    \end{minipage}
\caption{\label{tab:patterns} Pattern QUBOs for the four different types of clauses of the AlgorithmQUBO~\cite{Zielinski2023}.}
\end{table}

The variables $K$, in all of the Pattern QUBOs in Tab.~\ref{tab:patterns} represent additional, so-called \emph{ancilla variables}. They have no representation in the corresponding 3-SAT problem but need to be added to correctly represent a 3-SAT clause as a QUBO optimization problem (see Ref. \cite{Zielinski2023} for further details). Note, that for each clause a new ancilla variable needs to be added. Thus if there are $m$ clauses in the 3-SAT problem, this transformation needs $m$ additional variables.

\subsection{CountTrue: A 3-SAT Transformation with HOBO to QUBO Reduction}\label{sec:CountTrue}
We now introduce a straightforward approach to cast 3-SAT problems into QUBO form which is based on quadratization~\cite{Biamonte2008, Boros2018, Anthony2017, Dattani2019, Verma2021, Crama2022}. The approach we follow is to penalize (per clause) variable configurations
which do not satisfy a clause. It can also be viewed as counting the ${\rm{TRUE}}$ values per clause (since clauses with one, two, or three ${\rm{TRUEs}}$
are satisfied) and assign energy zero if the number of ${\rm{TRUEs}}$ is larger than zero and a positive energy value if there is no ${\rm{TRUE}}$ in the clause. This way, a solution with zero energy satisfies the 3-SAT problem, whereas a solution with energy larger than zero does not satisfy the problem. If we take a
clause $(l_1 x_{1} \vee l_2 x_{2} \vee l_3 x_{3})$ with $x_1, x_2, x_3 \in \{0,1\}$ and $l_1, l_2, l_3 \in \{\neg \neg\ , \neg\}$, we would need to construct a binary polynomial which penalizes configurations where $l_1 x_{1}=l_2 x_{2}=l_3 x_{3}=0$.
In general a polynomial with the aforementioned properties is the following
\begin{align}\label{eqn:hobo1}
	H_{\rm{CT}}(\vec{x}) = - \sum_{c=0}^{m-1} & \left(\sum_{(x,\chi_{x}) \in \mathcal{I}_{c}}\left(\frac{1 - \chi_{x}}{2} + \chi_{x} x\right) - 1\right) \\ \nonumber
						& \cdot \left(\sum_{(x,\chi_{x}) \in \mathcal{I}_{c}} \left(\frac{1 - \chi_{x}}{2} + \chi_{x} x\right) - 2\right) \\ \nonumber
						& \cdot \left(\sum_{(x,\chi_{x}) \in \mathcal{I}_{c}} \left(\frac{1 - \chi_{x}}{2} + \chi_{x} x\right) - 3\right) \, ,
\end{align}
where $\mathcal{I}_{c}$ is the set of bits $x$ and corresponding signs $\chi_{x}$ of literals in clause $c$. In particular, $\chi_{x}$ is the sign of the literal
in the clause, i.e., $\chi_{x}=1$ if $x$ is in the clause and $\chi_{x}=-1$ if $\neg x$ is in the clause. Unfortunately, $H_{\rm{CT}}$ is a higher order binary
polynomial (HOBO) because it contains cubic terms. A reduction to QUBO is straight forward by replacing a product of two bits by a new variable (e.g., $y\in \{0,1\}$)
and coupling the value of the new variable to the value of the product by a penalty polynomial. The penalty polynomial then needs to attain the value zero iff the new
variable has the same value as the product of the two bits. For example, in the product of three variables $x_{1} x_{2} x_{3}$ we replace $x_{2} x_{3}$ by $y$
where $y$ is a new bit variable. To enforce that $y=1$ iff $x_{1} x_{2}=1$ we need to add an additional penalty term of the form $x_{1} x_{2} - 2 x_{1} y - 2 x_{2} y + 3 y$.
It is worth pointing out that compared to the other transformations, CountTrue uses the minimal amount of auxiliary bits since they can often be reused in several clauses. See
also Fig.~\ref{fig:qubo_stats} for an overview of the bits needed for each of the transformations.

\subsection{Digital Annealer}
Besides Quantum Annealing and variational algorithms on gate-based quantum computers, Fujitsu's Digital Annealer Unit (DAU) is a very promising and
powerful technology to solve large and complex combinatorial optimization problems in form of QUBOs. The DAU is a custom application-specific integrated
circuit (ASIC) hardware architecture realized using conventional CMOS technology. A Markov Chain Monte Carlo (MCMC) algorithm forms the basis of the
search algorithm. By implementing the MCMC algorithm on hardware, the search for low energy solutions to an optimization problem can drastically be
enhanced. Specifically, using techniques such as parallel search, a dynamic offset energy to escape from local minima, and parallel tempering speeds up
the search. 
With version 2 of the DAU (DA V2)  problem instances with up to 8192 fully connected variables and an integer precision of 16 $\rm{bit}$ for the QUBO values
$Q_{ij}$ can be treated. If the integer precision needs to be increased to 64 $\rm{bit}$, the maximal problem size is reduced to 4096. In comparison to quantum
annealers, no special environment such as cryogenics, vacuum, etc. is needed. Another limiting factor present in current quantum annealers is the so-called
minor embedding which is absent in Digital Annealer due to the all-to-all connectivity of decision variables. In addition, problem sizes which are of practical
relevance and go beyond many toy model considerations when using quantum computing can be solved with Digital Annealer.

QUBO implementation and addressing the solver are done using Futjitsu's proprietary software development kit (SDK) called \verb|dadk|~\cite{datut}. Further details on the Digital Annealer, the SDK, solver parameter, and use cases can also be found in Ref.~\cite{datut}.

\section{Results}\label{sec:results}
In the following, we introduce and characterize the 3-SAT instances and furthermore present results in terms of percentage of solved 3-SAT instances and percentage of correct solutions using DA V2.

\subsection{3-SAT Instances}
\begin{figure}[!ht]
	\centering
	\includegraphics[width=0.99\columnwidth]{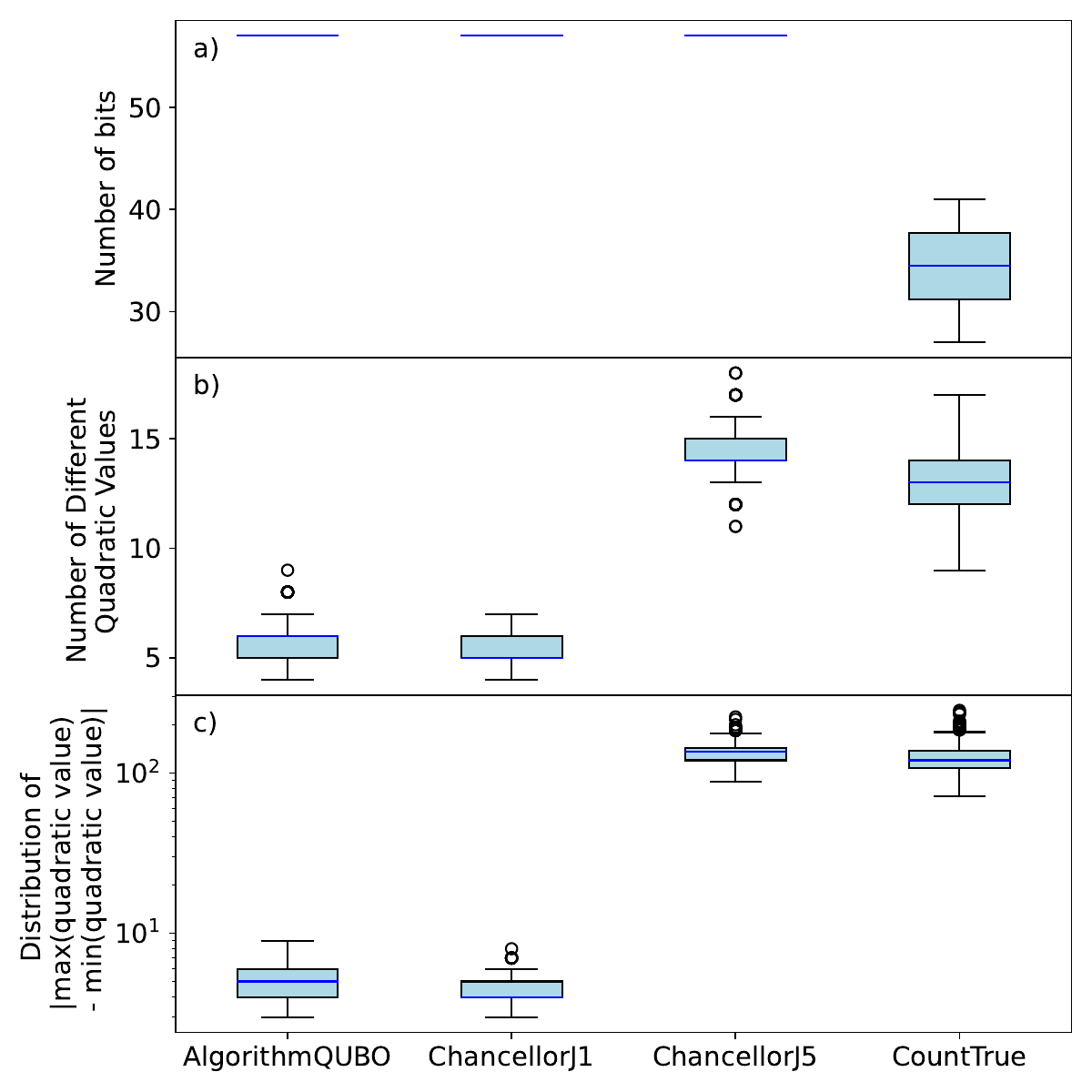}
	\caption{\label{fig:qubo_stats} QUBO matrix properties for each of the transformations. The plots show a boxplot characterizing all 1000 different
	3-SAT instances. a) Number of bits needed to represent the 3-SAT instance as a QUBO. Here, CountTrue varies in the number of bits which is due to
	the fact that we reuse auxiliary variables. b) Number	of different quadratic values of the QUBO matrix. c) Size of the value range of quadratic
	values, i.e., the absolute value of difference largest quadratic value minus smallest quadratic value of the QUBO matrix.
}
\end{figure}

We use the same 3-SAT instances with $m=46$ clauses and $n=11$ variables as in Ref.~\cite{Zielinski2023} to evaluate the performance of the different
transformations. The instances were randomly generated drawing from a uniform distribution. With $m/n=46/11\approx4.18$ the ratio is chosen close the 3-SAT phase transition which makes them potentially hard to solve~\cite{Gent1995}. In total we investigate 1000 different 3-SAT instances. As also pointed out in Ref.~\cite{Zielinski2023}, it proofs insightful to inspect properties
of the QUBO matrix for each of the transformations, namely the number of bits needed, the number of different coefficient values of quadratic terms, and the distribution of the
values range of the coefficient values of quadratic terms. We show these properties in Fig.~\ref{fig:qubo_stats}. We see that CountTrue needs the lowest amount of bits
which can be explained by the fact that we reuse auxiliary variables (for some instances auxiliary variables can be reused more often for other not).
From Fig.~\ref{fig:qubo_stats} we also see that ChancellorJ5 and CountTrue, as well as AlgorithmQUBO and ChancellorJ1 show similar structure in their QUBO matrices.

Furthermore, using the same instances as in Ref.~\cite{Zielinski2023} allows us to compare the results obtained in Ref.~\cite{Zielinski2023} with a D-Wave quantum annealer to the ones we obtain using Fujitsu's Digital Annealer Version 2.

\subsection{Experiment - Settings}
We use Digital Annealer V2 to solve each of the 1000 3-SAT instances 100 times. We make use of an automatic scaling which scales the entries of
the QUBO matrix in a way that the largest value of the QUBO matrix is less than $2^{p-1}$ where $p=64$ is the integer precision of coefficients in the QUBO matrix.
Furthermore, we use an automatic determination of the annealing temperatures, i.e., the start and end temperature as well as the dynamical
offset energy; details can be found in Ref.~\cite{datut} and App.~\ref{sec:app_temp}. We use DA V2 in annealing mode, not in parallel tempering mode.
For each transformation we vary the number of iterations of DA V2 from $10^3$ to $10^8$. In the case of the CountTrue transformation, we need
to reduce a binary polynomial of third degree to second degree.
This is conveniently done using the \verb|reduce_higher_degree| method of the \verb|BinPol| class in the \verb|dadk| SDK~\cite{datut}.

\subsection{Experiment - Results}
\begin{figure}[!ht]
	\centering
	\includegraphics[width=0.99\columnwidth]{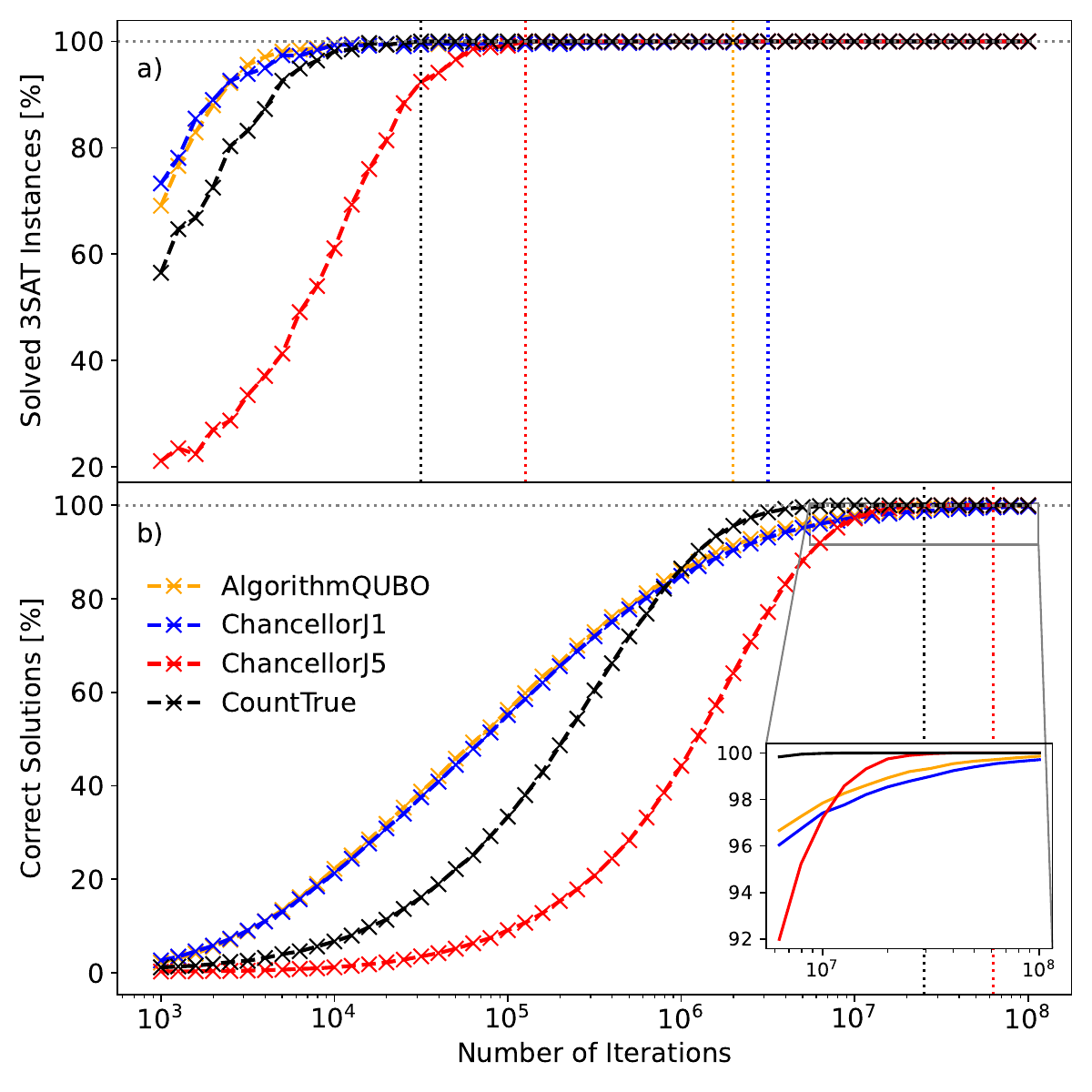}
	\caption{\label{fig:results} a) The percentage of solved 3-SAT instances increases with increasing number of iterations for all transformations.
	Vertical lines indicate the minimal number of iterations to reach $100\%$ correct solutions. b) The percentage of correct solutions also increases
	with the number of iterations. Two groups are forming among the transformations, showing similar behavior, namely, AlgorithmQUBO / ChancellorJ1 
	and CountTrue / ChancellorJ5. This grouping is also reflected in the properties of the QUBO matrix, see Fig.~\ref{fig:qubo_stats}. CountTrue shows
	the best performance which we attribute to the smaller amount of bits needed to represent the 3-SAT instances as QUBO.
}
\end{figure}

Figure~\ref{fig:results} shows the results of the experiment. In Fig.~\ref{fig:results} a) we show the percentage of solved 3-SAT instances and in
Fig.~\ref{fig:results} b) we show the percentage of overall correctly solved 3-SAT instances, i.e., out of $100 \times 1000$ experiments for each
fixed iteration number and all transformations.
We see that with increasing number of iterations, the percentage of solved 3-SAT instances and the percentage of correct solutions also increase.

Furthermore, we see that with all transformations it is possible to solve all 100 different 3-SAT instances [the vertical dashed colored lines in
Fig.~\ref{fig:results} a) indicate the minimal iteration number where $100\%$ of the 3-SAT instances are solved]. However, the probability for finding the
correct solution is (in most cases) still less than $100\%$ at this point, but for some transformations increases up to $100\%$ at higher iterations, cf. Fig.~\ref{fig:results} b).

In addition, we see from Fig.~\ref{fig:results} b) that for less iterations AlgorithmQUBO and ChancellorJ1 lead to a substantial amount of
correct solutions compared to CountTrue and ChancellorJ5. However, there is a crossover: with increasing number of iterations CountTrue and ChancellorJ5
outperform the other two transformations [also see inset of Fig.~\ref{fig:results} b)]. Actually, CountTrue and ChancellorJ5 are the only two transformations
reaching $100\%$ correct solutions [indicated by the vertical dashed colored lines in Fig.~\ref{fig:results} b)] at all.

The similar behavior of CountTrue and ChancellorJ5 on the one hand and AlgorithmQUBO and ChancellorJ1 on the other hand, is also reflected in
their similar properties of the QUBO matrix, see Fig.~\ref{fig:qubo_stats}. However, there are differences between CountTrue and ChancellorJ5:
CountTrue shows an earlier crossover behavior and reaches $100\%$ correct solution before ChancellorJ5. This makes CountTrue perform better
than ChancellorJ5 for large numbers of iterations. We attribute this fact to the smaller number of bits needed for CountTrue than for ChancellorJ5.

In Ref.~\cite{Zielinski2023} the D-Wave Advantage\_system4.1 was used to solve the same problem instances and the transformations ChancellorJ1
and ChancelorJ5 were also investigated. Comparing our results which we obtained with DA V2, we see that already with as few as $10^4$ iterations
DA V2 achieves better results than the D-Wave quantum annealer. With increasing number of iterations, DA V2 drastically outperforms the D-Wave
quantum annealer.
\begin{table}[!ht]
\centering
\begin{tabularx}{\columnwidth}{lc|ccc|cc}
\toprule
& \multicolumn{3}{c}{Solved 3-SAT instances $[\%]$} & \multicolumn{3}{c}{Correct solutions $[\%]$} \\
\cmidrule(lr){2-4}\cmidrule(lr){5-7}
& \multicolumn{1}{c}{D-Wave} & \multicolumn{2}{|c}{DA V2 \qquad \qquad} & \multicolumn{1}{c}{D-Wave} & \multicolumn{2}{|c}{DA V2} \\
DA iterations & \multicolumn{1}{c}{} & \multicolumn{1}{|c}{$10^4$} &  \multicolumn{1}{c}{$10^8$} & \multicolumn{1}{c}{} &  \multicolumn{1}{|c}{$10^4$} & \multicolumn{1}{c}{$10^8$} \\
\cmidrule(lr){2-4}\cmidrule(lr){5-7}
ChancellorJ1	& 91.0 & {\bf 99.3} & {\bf 100.0} & 8.71 & {\bf 21.27} & {\bf 99.7} \\
ChancellorJ5	& 58.0 & {\bf 61.1} & {\bf 100.0} & 0.48 & {\bf 1.16}   & {\bf 100.0} \\ \bottomrule
\end{tabularx}
\caption{
\label{tab:dwave_da} Comparison between D-Wave quantum annealer and DA V2. For DA V2 we show the results for $10^4$ and $10^8$ iterations. Numbers for D-Wave
are taken from Refs.~\cite{Zielinski2023} where the results were obtained using default parameters. Also, from Fig.~\ref{fig:results} it is obvious that with increasing
iterations, the solution quality of DA V2 is much better than using a D-Wave quantum annealer.
}
\end{table}
The D-Wave results were obtained using the default parameter settings with an anneal time of $20 \, {\rm{\mu s}}$. For DA V2, $10^4$ iterations
result in $\approx 16 \, {\rm{ms}}$ anneal time on the DA V2 chip. For $10^6$ iterations this time increases to $\approx 142 \, {\rm{ms}}$ anneal time
and for $10^8$ iterations the anneal time on the DA V2 chip is $\approx 12.8 \, {\rm{s}}$. Times are for obtaining all $100$ solutions. Since the anneal time on the quantum annealer is limited
(on the hardware side) by the qubit coherence time, a comparison on the time to solution level is hard.

\section{Analysis of Transformations and their Performances}\label{sec:analysis}
In order to better understand the differences in performance of the transformations with regards to the solution quality, especially the percentage of correct solutions, it would be very instructive to know more about the energy landscape of the QUBO formulation resulting from the different transformations.

Although the QUBOs are very small ($<100$ bits), exact diagonalization is out of reach to learn more about eigenvalues and eigenstates. This is due to the fact
that the corresponding Hamiltonian matrix grows exponentially with the number of bits needed to represent the 3-SAT instance as a QUBO. Therefore, we study small 3-SAT instances
and diagonalize the corresponding Ising Hamiltonian such that we are able to investigate the eigensystem of the Hamiltonian matrix. Here, we take ten
randomly created 3-SAT instances with $m=20$ clauses and $n = 5$ variables. The properties of the QUBO matrices are shown in Fig.~\ref{fig:results_ed} a)-b).
As in the case of the large instances (see Fig.~\ref{fig:qubo_stats}), also for the small instances, we encounter two groups, AlgorithmQUBO / ChancellorJ1 and CountTrue / ChancellorJ5 which show similar QUBO properties. 

In the next sections, we analyze the small instances exactly, explore correlations to the solution quality, and extrapolate the gained insights to the larger 3-SAT instances.

\subsection{Exact Diagonalization}
Diagonalization of the QUBO matrix without constraining to binary variables would lead to solutions which are not meaningful to the actual problem. Therefore,
we apply the transformation $F(\vec{x}):=\bigotimes_{i=0}^{n-1} f_i (x_i)$, with
\begin{align}\label{eqn:ed11}
	f_i: \{0,1\}\rightarrow \mathbb{R}^2,\quad u\mapsto \left(\begin{matrix}u\\ 1-u\end{matrix}\right),
\end{align}
to the binary vector $\vec{x}\in \{0,1\}^n$. We then examine the QUBO problem in the vector space $\mathbb{R}^{2^{n}}$. In particular, we elevate
each binary variable $x_{i}$ occurring in the QUBO to a matrix by mapping $x_{i} \mapsto (I_{i}+Z_{i})/2$ where $I$ is the $2 \times 2$ identity matrix
and $Z={\rm{diag}}(1,-1)$ is the Pauli-$z$ matrix. Identity matrices have to be inserted appropriately when transforming. For example, in a QUBO with
$n$ binary variables $x_{i}$, $i=0,\ldots,n-1$, linear terms $\sim x_{i}$ are mapped by
\begin{align}\label{eqn:qi_map_x}
	x_{i} \mapsto \bigotimes_{0<\alpha < i} I_{\alpha} \otimes \frac{I_{i}+Z_{i}}{2} \otimes \bigotimes_{i<\beta \leq n-1} I_{\beta} \, ,
\end{align}
and quadratic terms $\sim x_{i} x_{j}$ are mapped in the following way
\begin{align}\label{eqn:qi_map_xx}
	x_{i} x_{j} \mapsto &\bigotimes_{0<\alpha < i} I_{\alpha} \otimes \frac{I_{i}+Z_{i}}{2} \otimes \\ \nonumber
				&\bigotimes_{i<\beta<j} I_{\beta} \otimes \frac{I_{j}+Z_{j}}{2} \otimes \\ \nonumber
				&\bigotimes_{j<\gamma \leq n-1} I_{\gamma} \, .
\end{align}
The total Hamiltonian matrix is then the sum of all transformed QUBO terms and is by construction diagonal.

For the small instances we construct the QUBO matrix as before, do the mapping according to Eqs.~(\ref{eqn:qi_map_x}) and (\ref{eqn:qi_map_xx}), and exactly diagonalize the
resulting Hamiltonian matrix. First, we have a closer look at the degeneracy of the smallest eigenvalue of the spectrum.
\begin{figure}[!ht]
	\centering
	\includegraphics[width=0.99\columnwidth]{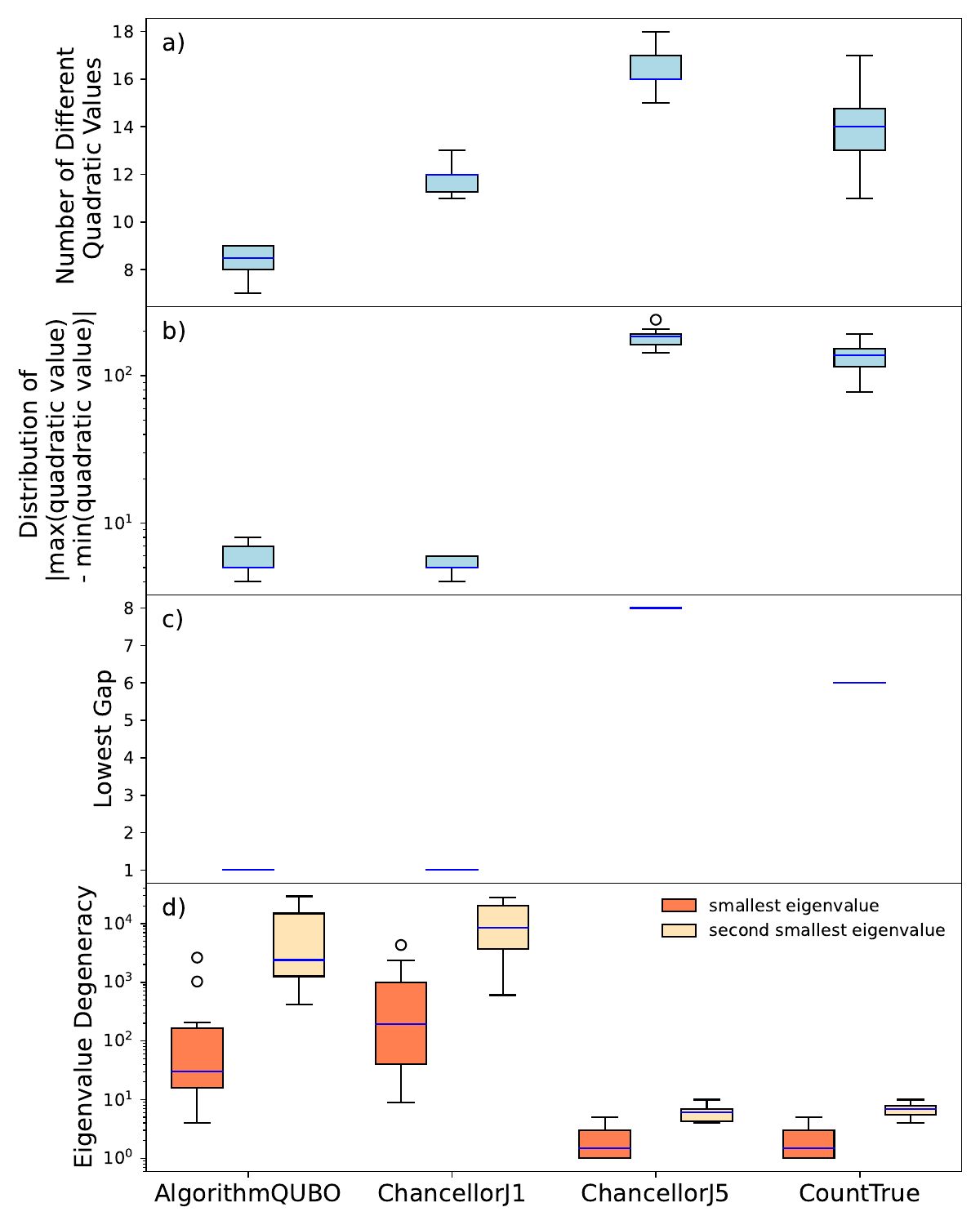}
	\caption{\label{fig:results_ed} QUBO matrix properties for each of the transformations for small 3-SAT instances $m=20$ and $n = 5$ .
	The plots show a boxplot characterizing all 10 different 3-SAT instances. a) Number of different quadratic values of the QUBO matrix.
	b) Size of the value range of quadratic values, i.e., the absolute value of difference between largest quadratic value and smallest quadratic
	value of the QUBO matrix. c) Energy gap between the smallest eigenvalue and the second smallest eigenvalue. d) Degeneracy of the
	smallest eigenvalue (orange) and second smallest eigenvalue (light orange). As in the case for the other instances (cf. Fig.~\ref{fig:qubo_stats})
	two groups AlgorithmQUBO / ChancellorJ1 and CountTrue / ChancellorJ5 can be distinguished.
}
\end{figure}
In Fig.~\ref{fig:results_ed} d) we show the degeneracy of the smallest and second smallest eigenvalue for each transformation. Again, we see two
groups, the first group AlgorithmQUBO / ChancellorJ1 shows a relatively high degeneracy (both for the smallest and second smallest eigenvalue),
in contrast to that the second group CountTrue / ChancellorJ5 shows a relatively low degeneracy. This grouping also exists for the energy gap between
the smallest and second smallest eigenvalues, shown in Fig.~\ref{fig:results_ed} c). Except for the CountTrue transformation, the search space is the same.

Since, the degeneracy of the smallest eigenvalue is directly related to the number of solutions of the problem instance, a high ground state degeneracy
(at constant size of the search space) leads to an increased probability of finding this minimum energy state. This explains the behavior of the transformations
shown in Fig.~\ref{fig:results}.
At already low number of iterations, the transformations AlgorithmQUBO / ChancellorJ1 yield quite good solutions, which is due to the fairly
large number of correct solutions (i.e., high degeneracy of the smallest eigenvalue) with respect to the size of the search space. Increasing the number of
iterations saturates the curves for AlgorithmQUBO and ChancellorJ1 in Fig.~\ref{fig:results} b).
For CountTrue / ChancellorJ5, however, it is harder to find a minimum energy solution with just a few iterations because the number of solutions is
very small compared to the size of the search space. Increasing the number of iterations increases the probability to find a minimum energy solutions for CountTrue / ChancellorJ5.

Moreover, the transformations AlgorithmQUBO / ChancellorJ1 also have a high degeneracy of the second lowest eigenvalue which are of low energy
[see the gap shown in Fig.~\ref{fig:results_ed} c)] but which do not satisfy the SAT instances. This fact might hinder the improvement in correct
solutions with increasing iterations, since many iterations might be spent on flipping back and forth between configurations corresponding to the
second lowest eigenvalue (which have the same energy value) rather than finding configurations with minimal energy value (i.e. correct solutions).

In contrast, the CountTrue / ChancellorJ5 transformations show a small number of correct solutions for small iterations which is due to the low
degeneracy of the lowest eigenvalue. However, they show good convergence towards the optimal solution with increasing iterations which is
due to the low degeneracy of the second lowest eigenvalue. In particular, for CountTrue / ChancellorJ5, there are fewer energetically semi-good but incorrect configurations (because these correspond to the second lowest eigenstate) in which the annealing process might get stuck.

From Fig.~\ref{fig:results} b), we see that CountTrue and ChancellorJ5 do not behave as similar as AlgorithmQUBO / ChancellorJ1. To be precise, the curve (c.f. Fig.~\ref{fig:results} b) black and red lines) for the CountTrue transformation is slightly shifted to the left in comparison to the curve for the ChancellorJ5 transformation. We
attribute this shift to the smaller search space of the CountTrue transformation which is due to the smaller number of bits needed.
 
To summarize, the insights from analyzing small instances exactly can be extrapolated to the large instances and provide a reasonable ground to
explain differences in the transformations on the solution of 3-SAT instances.

\subsection{Hamming Distance}
\begin{figure}[!ht]
	\centering
	\includegraphics[width=0.99\columnwidth]{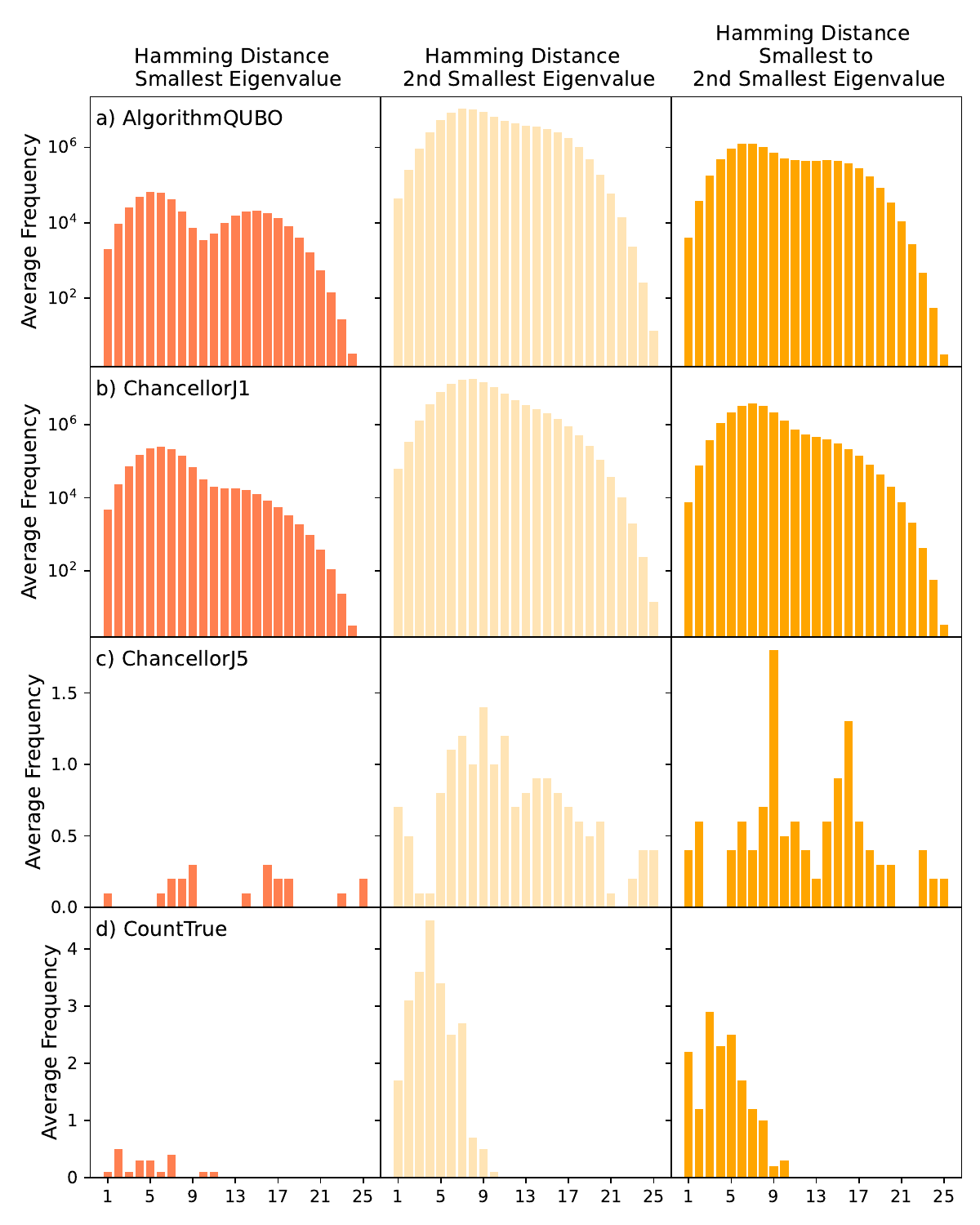}
	\caption{\label{fig:results_hd} Average frequency of the Hamming distance among eigenstate configurations corresponding to the smallest eigenvalues,
	the eigenstate configurations corresponding to the second smallest eigenvalues, and the eigenstate configurations corresponding to the smallest and the second smallest
	eigenvalues (left panels to right panels) for each transformation [a) to d)]. We see a grouping of AlgorithmQUBO / ChancellorJ1 and
	CountTrue / ChancellorJ5 explaining their similar influence on the solution. However, CountTrue and ChancellorJ5 show slightly different statistics
	which is due to their different search spaces.
}
\end{figure}

To further strengthen the argument we investigate the difference of the eigenstate configurations (i.e., solution configurations corresponding to the eigenstates of minimal eigenvalue) in terms of the Hamming distance
between eigenstate configurations. We show the statistics of the Hamming distance for the eigenstate configurations corresponding to the smallest eigenvalues, the eigenstate configurations corresponding to the second smallest eigenvalues, and the eigenstate configurations corresponding to the smallest and the second smallest eigenvalues in
Fig.~\ref{fig:results_hd}.
Again, we see a grouping of AlgorithmQUBO / ChancellorJ1 and CountTrue / ChancellorJ5. This time however with subtle differences between the
CountTrue and ChancellorJ5 transformation.

As a first observation, we see that the average frequency of Hamming distance between eigenstate configurations of smallest eigenvalues and also of the second smallest eigenvalues
is higher for the ChancellorJ1 transformation compared to the AlgorithmQUBO transformation. In particular, AlgorithmQUBO and ChancellorJ1 show
a peak at Hamming distance eight between different configurations corresponding to the second smallest eigenvalue. The peak is about 1.5 times
higher for ChancellorJ1. Extrapolating to the larger 3-SAT instance and the result presented in Fig.~\ref{fig:results}, this explains the slightly slower growth of ChancellorJ1 in Fig.~\ref{fig:results} which can then be associated to spending many iterations on flipping between configurations
corresponding to the second smallest eigenvalue.

Concerning the CountTrue and ChancellorJ5 transformation, we see a similar degeneracy for smallest and second smallest eigenvalues. However,
there are slight differences in the average Hamming distance frequencies between configurations corresponding to smallest, second smallest, and
smallest to second smallest eigenvalues.
The average frequency for Hamming distance between configurations corresponding to the smallest eigenvalue is always lower than one for both
transformations. This is due to the fact that for most formulas there are only a single or very few valid solutions for CountTrue and ChancellorJ5.
This explains the similar behavior of both transformations for small iteration numbers with a slight advantage for CountTrue probably due to the
smaller search space. 
Concerning the second smallest eigenvalues and corresponding configurations, the average frequency for small Hamming distances is slightly higher
for CountTrue than for ChancellorJ5. In principle, we would therefore expect a slower increase in correct solutions with respect to iteration numbers
for CountTrue. We account the opposite behavior in Fig.~\ref{fig:results} for low iteration numbers to the smaller search space of CountTrue which
seems to compensate the expected slower convergence due to smaller Hamming distances of configurations corresponding to the second smallest
eigenvalue. At the same time, the degeneracy for the second smallest eigenvalue is very low both for CountTrue and ChancellorJ5 which also mitigates
the effect of Hamming distances of configurations corresponding to the second smallest eigenvalue. However, the convergence above
$80\%$ solved instances / correct solutions is indeed slower for CountTrue compared to ChancellorJ5 as seen in Fig.~\ref{fig:results},
which indicates that there exist instances for which the smaller Hamming distance between configurations
corresponding to the second smallest eigenvalue is problematic.

\section{Conclusion}\label{sec:conclusion}
To summarize, we have studied different 3-SAT to QUBO transformations and their influence on the solution quality when solving 3SAT instances with Fujitsu's Digital Annealer. In addition to the previously studied 3-SAT to QUBO transformations, we have introduced a novel transformation, CountTrue.
We have seen that different transformations have different impact on the solution quality. We explained the different behavior by means of exact diagonalization of small 3-SAT instances and extrapolated the insights to larger instances. This analysis also explains the good performance of our novel CountTrue transformation.
Furthermore, we have compared our results to results obtained with D-Wave's quantum annealer Advantage 4.1. This comparison shows a clear advantage of Digital Annealer over the quantum annealer in terms of solution quality. It is worth mentioning that the performance ranking of different transformations is the same on the quantum annealer and on the Digital Annealer. 

With these insights, we plan to investigate tailoring of ground state degeneracies to enhance the solution quality. In addition, we plan to investigate larger 3-SAT instances.

\textit{Acknowledgements}.-- We would like to acknowledge stimulating discussions with Markus Kirsch, and valuable comments by Jan Budich and Andreas Rohnfelder.

\begin{widetext}
\appendix

\section{Automatic Temperature Determination}\label{sec:app_temp}
A well adjusted search process for the Digital Annealer typically has two phases. In the first phase the temperature is sufficiently high to instantly escape 
from local minima. In this phase the random walk has progress in every step. In the second phase temperatures should be low enough to produce waiting
cycles before escaping from a local minimum. Both phases and the transition point from the first to the second phase can be controlled by the start and end
temperature.
To estimate the dependency of the flip probability at a local minimum from the temperature let $X$ be a minimum state and $X^i$ be the state for which
the bit with index $i$ is flipped:
\begin{align}\label{eqn:app1}
    x^i_j=
    \begin{cases}
      x_j & \text{for } i \ne j \\
      1 - x_j & \text{for } i = j \, .
    \end{cases}
\end{align}

Let $T$ be the temperature and $E(X)$ and $E(X^i)$ the energies of the states. Since $X$ is a local minimum we can calculate the flip probability as follows:
\begin{align}\label{eqn:app2}
	p_{\rm{flip}} = 1 - \prod_{i=0}^{N-1} \left(1 - e^{- \frac {E(X_i)-E(X)} T} \right) \, .
\end{align}
Let $\Delta E(X)$ be a random variable defined on the set of minima of $E$ as the mean of energy differences for all bit flips
\begin{align}\label{eqn:app3}
	\Delta E(X) = \frac 1 N \sum_{i=0}^{N-1} \left( E(X_i)-E(X) \right) \, .
\end{align}
As rough estimation we next replace all energy differences in Eq.~(\ref{eqn:app3}) by the expectation value of $\Delta E(X)$
\begin{align}\label{eqn:app4}
	p_{\rm{flip}} \approx 1 - \prod_{i=0}^{N-1} \left(1 - e^{- \frac {\mathbb{E}(\Delta E(X))} T} \right)
\end{align}
which leads to an estimation for a suitable temperature
\begin{align}\label{eqn:app5}
	T \approx - \frac {\mathbb{E}(\Delta E(X))} {\ln(1 - \sqrt[N]{1-p_{\rm{flip}}})} \, .
\end{align}

\noindent{{\bf{\emph{Start Temperature}}}} \\
In the beginning, waiting cycles in local minima should be avoided. Therefore, we define a high flip probability, e.g., $p_{\rm{start}}=0.99$. According to
Eq.~(\ref{eqn:app5}) we get for the start temperature
\begin{align}\label{eqn:app6}
	T_{\rm{start}} = - \frac {\mathbb{E}(\Delta E(X))} {\ln(1 - \sqrt[N]{1-p_{\rm{start}}})} .
\end{align}

\noindent{{\bf{\emph{Transition Temperature}}}} \\
For the transition point between the first and the second phase waiting cycles in local minima should become more likely. Therefore, we define a
lower flip probability, e.g., $p_{\rm{trans}}=0.5$. According to Eq.~(\ref{eqn:app5}) we get for the transition temperature
\begin{align}\label{eqn:app7}
	T_{\rm{trans}} = - \frac {\mathbb{E}(\Delta E(X))} {\ln(1 - \sqrt[N]{1-p_{\rm{trans}}})} \, .
\end{align}

\noindent{{\bf{\emph{End Temperature}}}} \\
Let $I$ be the total number of iterations. Let the transition point be defined by a part of iterations $0 < \nu \le 1$. Let $d:=1- \rm{decay}$ be the
factor multiplied for exponential temperature model in every step to the previous temperature. Then we have $T_{\rm{start}}d^{I} = T_{\rm{end}}$
and $T_{start}d^{\nu I} = T_{trans}$. The first equations gives $d^I = T_{start}^{-1} T_{end}$ and using this to replace $d^I$ in the
second equation, we get $T_{\rm{start}}  T_{\rm{start}}^{-\nu} T_{\rm{end}}^{\nu} = T_{\rm{trans}}$. Therefore
$T_{\rm{end}} = T_{\rm{start}}^{1- \frac 1 \nu} T_{\rm{trans}}^{- \frac 1 \nu}$ and with Eq.~(\ref{eqn:app7}) we get
\begin{align}\label{eqn:app8}
	T_{\rm{end}} = T_{\rm{start}}^{1- \frac 1 \nu} \left( - \frac {\mathbb{E}(\Delta E(X))} {\ln(1 - \sqrt[N]{1-p_{\rm{trans}}})} \right)^{- \frac 1 \nu} \, .
\end{align}

\noindent{{\bf{\emph{Offset Parameter}}}} \\
Finally we have to determine the dynamic energy offset that should help to escape faster from local minima. To climb safely upwards from a local
minimum, we add only a fraction of the expected depth of the minimum in every step. If we want to reach the rim in, e.g., $k:=10$ steps, then we define 
\begin{align}\label{eqn:app9}
	E_{\rm{dyn\_off}} :=  \frac {\mathbb{E}(\Delta E(X))} k \, .
\end{align}
%

\end{widetext}


\end{document}